\documentclass[aps,prd,showpacs,nofootinbib,floats,floatfix,preprintnumbers,groupedaddres]{revtex4}
\usepackage{comment}
\usepackage{graphicx}
\usepackage{amsmath, mathrsfs}
\usepackage{float}
\usepackage{subfigure}
\usepackage[usenames]{color}
\usepackage{amssymb}
\usepackage{bbm}
\usepackage{csquotes}
\usepackage{tensor}

\usepackage{tikz}
\usetikzlibrary{positioning,decorations.pathmorphing}

\usepackage{epstopdf}
\newcommand{\bea}{\begin{aligned}}
\newcommand{\eea}{\end{aligned}}
\newcommand{\beq}{\begin{equation}}
\newcommand{\eeq}{\end{equation}}

\newcommand{\bse}{\begin{subequations}}
\newcommand{\ese}{\end{subequations}}

\usepackage{hyperref}
\hypersetup{
     colorlinks   = true,
     citecolor    = red,
     linkcolor    = blue,
     urlcolor     = blue,
}

\renewcommand{\v}[1]{\ensuremath{\mathbf{#1}}} 
\newcommand{\bmm}{\begin{multline}}
\newcommand{\emm}{\end{multline}}

\def\={\stackrel{\Delta}{=}}
\def\ula#1{\underleftarrow{#1}}
\def\lie{\pounds}
\def\l{\ell}

\begin{document}
\title{Laws of black hole mechanics in the Einstein-Gauss-Bonnet theory}
\author{Ayan Chatterjee}
\email{E-mail: ayan.theory@gmail.com}
\affiliation{Department of Physics \& Astronomical Science, Central University of Himachal Pradesh, Dharamshala-
176215, India.}
\author{Sahil Devdutt}
\email{E-mail:devduttsahil@gmail.com }
\affiliation{Department of Physics \& Astronomical Science, Central University of Himachal Pradesh, Dharamshala-
176215, India.}

\author{Avirup Ghosh}
\email{E-mail: avirup.avi@gmail.com}
\affiliation{Department of Physics, Kulti College, Paschim Bardhaman, West Bengal- 713343.}
\pagenumbering{arabic}
\renewcommand{\thesection}{\arabic{section}}
\begin{abstract}
 We extend the isolated horizon formalism
to include rotating black holes arising in five dimensional Einstein- Gauss- Bonnet (EGB) theory of gravity, and derive the laws of black hole mechanics. This result allows us to show that the first law of black hole mechanics is modified, due to the Gauss- Bonnet term, so as to include corrections to (i) the area of horizon cross- sections and, to (ii) the expression of horizon angular momentum. Once these modifications are included, the Hamiltonian generates an evolution on the space of solutions of the EGB theory admitting isolated horizon as an internal boundary, the consequence of which is the first law of black hole mechanics. These boundary conditions may help in the search for exact solutions describing rotating black holes in this theory.  
\end{abstract}

\maketitle


The Isolated Horizon (IH) formalism of black holes is a quasilocal description of black hole horizons in equilibrium \cite{Ashtekar:1998sp, Ashtekar:2000sz,Ashtekar:2004cn} (a recent review is in \cite{Ashtekar:2025wnu}). The definition imposes conditions on the geometric and matter fields residing on the horizon, but does not enforce restrictions on the nature of spacetime or their symmetries at asymptotic infinity, or even near the horizon. Therefore, this quasilocal definition is able to accommodate a large class of black hole spacetimes, many of which may even admit radiation close to the horizon provided none of these cross the horizon. More precisely, the IH formalism requires that there exist a null Killing vector (on the horizon only) representing the fact that the horizon is time independent.  Naturally, this definition subsumes other definitions of black hole horizon like the Killing or the Event Horizon, which require a timelike Killing vector field near the horizon, or even in the full spacetime. In fact, the IH formalism has provided a useful quasilocal framework to address many questions in theoretical and numerical aspects of relativity. During the last several years, it has been possible to use this definition to construct the classical phase space in general relativity admitting IH as an internal boundary \cite{Ashtekar:1999yj,Ashtekar:2000hw,Ashtekar:1999wa,Chatterjee:2008if}. This phase space has led to a better understanding of the zeroth and the first laws of black hole mechanics and at the same time, has furnished a mechanism to develop a microscopic understanding 
of black hole entropy and its quantum correction from horizon microstates \cite{Ashtekar:1997yu,Ashtekar:2000eq,Kaul:2000kf,Meissner:2004ju,Domagala:2004jt,Ghosh:2004wq,Chatterjee:2020iuf}. Although the IH formalism has been immensely successful in general relativity, only little progress has been made to extend them to other theories of gravity (a careful study of the action and topological terms in the context of the IH formalism has been carried out in \cite{Corichi:2016zac,Corichi:2016eoe,Corichi:2018drc,Corichi:2025sbm,Perez:2017cmj}). The laws of mechanics for an IH in the scalar- tensor theory of gravity \cite{Ashtekar:2003jh}, and for non- rotating black holes in the Einstein- Gauss- Bonnet (EGB) theory of gravity are known \cite{Liko:2007th}. The phase space of rotating black holes in the EGB theory remains unexplored, probably due to lack of exact solutions. To address this gap, in this paper, we shall look into an extension of the IH formalism for rotating black holes in the EGB theory in five dimensions. We shall show that it is possible to extract expressions for area and angular momentum of these black holes in terms of the Ricci rotation coefficients. \\

Let $(\mathcal{M}, g_{ab}, \nabla)$ be a $5$- dimensional spacetime with signature 
$(-++++)$, metric $g_{ab}$, and a metric compatible covariant derivative operator $\nabla$. A null hypersurface $\Delta$ in $\mathcal{M}$, generated by a future directed null normal $\ell^{a}$, and endowed with a degenerate metric $q_{ab}\=g_{\ula{ab}}$ (the symbol $\=$ is used to denote equalities which are true only on $\Delta$, and arrow under the indices shows that indices are pulled back to $\Delta$), is said to be an weakly isolated horizon (WIH) if the following conditions hold:
\begin{enumerate}
    \item $\Delta$ is topologically $\mathbb S\times \mathbb{R}$, where $\mathbb S$ is a closed three dimensional manifold. 
    \item The expansion scalar of the null normal $\ell^{a}$, given by $\theta_{(\ell)}=q^{ab}\, \nabla_{a}\ell_{b}$ is vanishing on $\Delta$, that is, $\theta_{(\ell)}\=0$.
    \item $-R_{a}{}^{b}\,\ell^{a}$ is future directed and null , and all field equations hold on $\Delta$.
    \item On $\Delta$, the connection on the normal bundle $\omega_{a}$, defined through $\nabla_{\ula{a}}\, \ell^{b}\=\omega_{a}\, \ell^{b}$ is such that $\lie_{\ell}\,\omega_{\ula{a}}\=0$ .
\end{enumerate}
The first three conditions given above are used to define 
a non- expanding horizons (NEH) in $4$ and lower dimensions. 
The condition on cross sectional topology is assumed, keeping in mind that black holes horizons in higher dimensions can have more complicated topologies. The vanishing of the expansion scalar $\theta_{(\ell)}$ indicates that no matter or gravitational field is allowed to cross the horizon. The third condition, on $R_{ab}$ is usually placed, in $4$ dimensions, on the energy- momentum tensor, $T_{ab}$, and then related to geometrical tensors through the Einstein equations. Here, since we shall be dealing with somewhat more complicated field equations, the conditions are imposed directly on the Ricci tensor $R_{ab}$. Also, quite naturally, we need that the matter and geometric field equations shall continue to hold true on the horizon $\Delta$. The fourth condition is a restriction placed on the connection $\omega_{a}$ corresponding to $\ell^{a}$, such that $\lie_{\ell}\, \omega\=0$. This condition is equivalent to requiring that the connection corresponding to the null normal is lie dragged,  $[\lie_{\ell}, \, \nabla_{\ula{a}}]\, \ell^{a}\=0$.  \\

To understand the consequences of the above-mentioned boundary conditions, the horizon geometry needs to be specified. The horizon $\Delta$ is foliated by spacelike, $v=$constant, $3$-dimensional manifold $\mathbb S$. We choose the two null normals to these surfaces as $\ell^{a}$ and $n^{a}$, with the one-form $n_{a}$ being such that $n_{a}=-(dv)_{a}$.  The spacelike vectors $m_{(i)}^a$ form the basis for the tangent bundle $T(\mathbb S)$. Let us introduce the following notation: (The geometrical setup is adopted from \cite{Ortaggio:2007eg,Pravda:2004ka}, but we have made modifications to suit the present problem)
\begin{align}\label{ortho1}
    m^{(0)}&=n & m^{(1)}&=\ell & m^{(i)}&= \text{spacelike covectors}\\ \label{ortho2}
    m_{(0)}&=-\ell & m_{(1)}&=-n &  m_{(i)}&= \text{spacelike vectors},
\end{align}
such that the following orthonormality conditions hold: $\ell\cdot n=-1,\, \ell\cdot m^{(i)}=0,\, n\cdot m^{(i)}=0,\, m^{(i)}\cdot m^{(j)}=\delta^{(ij)}$. Note here we have introduced a common label $m^{(a)}$ for all the frame vectors. the indices $a,b,c$ run from $0$ to $4$ and the indices $i,j,k$ run from $2$ to $4$, therefore represent only spacelike vectors $m_{(i)}$. Let us define the following Ricci rotation coefficients:
\begin{eqnarray}\label{Ricci_coeff}
    \nabla_{b} \,\ell_{a} &=:& L_{cd}\, m_a^{(c)}m_b^{(d)}\label{100}\\
    \nabla_{b} \, n_{a}& =:& N_{cd}\, m_a^{(c)}m_b^{(d)}\\
    \nabla_{b} \, m_{a}^{(i)}& =:&  \stackrel{i}{M}_{cd}\, m_a^{(c)}m_b^{(d)}
\end{eqnarray}
The orthonormality conditions in equation eqn. \eqref{ortho1}, and $\eqref{ortho2}$ present us with the following constraints on the Ricci rotation coefficients:
\begin{eqnarray}\label{Ricci_cond}
   L_{i a} - \stackrel{i}{M_{0a}}&=0, ~~  & N_{0a}+L_{1a}=0, ~~~ L_{0a} = 0,\\\label{Ricci_cond_1}
    N_{i a} - \stackrel{i}{M_{1a}}&=0, ~~~ & \stackrel{j}{M_{ia}}+\stackrel{i}{M_{ja}} =0,~~~  \stackrel{i}{M_{ia}} =0, ~~~   N_{1a}=0. \label{Ricci_cond_2}
\end{eqnarray}
Further constraints on Ricci rotation coefficients arise from the isolated horizon boundary conditions: First, a geodetic $\ell^a$ implies $L_{i0}\=0$. This is obtained as follows: From equation eqn.\eqref{Ricci_coeff}, it is clear that $\ell^b\nabla_b \ell_{a}=L_{c0} m_a^{(c)}=L_{10}\ell^{a}+L_{i0}m^{(i)}_a$, and therefore, since the geodetic condition requires that the right side be proportional to $\ell^{a}$, the result follows. Using $L_{0a}=0$ and $L_{i0}=0$ we can write:
\begin{equation}
    \nabla_{\underleftarrow{b}}\ell^{a} \=\left[L_{10} n_b + L_{1i} m_b^{(i)}\right]\,\ell^{a} + L_{ij}m_b^{(j)}m^{(i)\,a}.
\end{equation}
    The matrix $L_{ij}$ can be decomposed in the symmetric (and tracefree) and the antisymmetric parts in the following manner: $L_{ij}=S_{ij}+A_{ij}$, where $S_{ij}=L_{(ij)}=\sigma_{ij}+(\theta_{(\ell)}/3)\delta_{ij}$ and $A_{ij}=L_{[ij]}=\omega_{ij}$ is the twist corresponding to the generator $\ell^{a}$. But since IH boundary is a twist-free, shear- free and expansion- free null surface, $L_{ij}\=0$. It further follows from eqns. \eqref{Ricci_cond}, and \eqref{Ricci_cond_1}, that the following quantities vanish identically on $\Delta$:\\
\begin{equation}
    L_{0a}\= L_{i0}\=L_{ij}\=~\stackrel{i}{M_{00}}~\=~\stackrel{i}{M_{0j}}~\=0. 
\end{equation}

These expressions allow us to evaluate the coefficents 
of the spin- connection $A_{IJ}$ on the horizon $\Delta$.
It may be argued that $A_{IJ}$ may be expanded in the internal basis as follows:
\begin{equation}\label{connection_expansion}
    \tensor{A}{_{\underleftarrow{a}}^{IJ}}\stackrel{\Delta}{=}-2 \omega_a \ell^{[I}n^{J]}+2 V^{(i)}_a \ell^{[I}m^{J]}_{(i)}+ W^{(ij)}_a m_{(i)}^{[I}m^{J]}_{(j)}.
\end{equation}
This may be obtained in the following manner: Let the
internal basis $[\,\ell^{I}, n^{I}, m_{(i)}^{I}\,]$ are fixed on $\Delta$ such that 
the derivative operator $\partial_{a}$
annihilates them, and therefore, $\nabla_{a}\ell^{I}={A_{a}}^{I}{}_{J}\, \ell^{J}$.
Now, rewriting the equation $\nabla_{\ula{a}}\ell^{b}\= \omega_{a} \ell^b$, and using that
\begin{equation}
    \nabla_{\underleftarrow{a}} \ell^{b}\= \omega_{a} \ell^b \= \nabla_{\underleftarrow{a}}(e^{b}_I \,\ell^I)=e^{b}_{I}\, A_{\ula{a}}^{I}{}_{J}\ell^{J},
\end{equation}
we obtain that the component of spin- connection should contain: $A^{IJ}=-2\,\omega_{a}\, \ell^{[I}\, n^{J]}$.
The covariant derivatives of the other vector fields
are also known:
\begin{eqnarray}\label{11}
    \nabla_{\underleftarrow{a}}n^b&\= &-\omega_a n^b +V^{(i)}_a m_{(i)}^{b}\\
\nabla_{\underleftarrow{a}}m^b_{(i)}&\= &V_a^{(i)}\ell^b-W_a^{(ij)}m^b_{(j)}, ~~~~~~~~~\, (i\ne j).
\end{eqnarray}
%
Further the choice that $n=-dv$ implies that $dn=0$, from which, we get $N_{[ij]}\=0$ and $N_{0i}\=N_{i0}$. It is useful to calculate 
the curvature $F_{IJ} =dA_{IJ}+A_{IK}\wedge\, A^{K}{}_{J}$ corresponding
to this connection $A_{IJ}$ in equation \eqref{connection_expansion}. First, note that the rotation one-form $\omega_{a}$ in the eqn. \eqref{connection_expansion} may be written as:
\begin{equation}\label{omega_expansion}
\omega_{a}\=-\kappa_{(\ell)} n_a + \Tilde{\omega}_{a},
\end{equation}
where $\Tilde{\omega}_{a}$ is pullback of the one form on the cross-sections, and if the horizon is non-rotating then $\Tilde{\omega}_a=0$.
Also the other two connection forms $V^{i}$ and $W^{ij}$ can be written as:
\begin{eqnarray}\label{W and V}
V^{(i)}_{\underleftarrow{b}}&\stackrel{\Delta}{=}& N_{i0}n_b+N_{ij}m^{(j)}_{b}\\ \nonumber
W^{(ij)}_{\underleftarrow{b}}&\stackrel{\Delta}{=}&-\stackrel{i}{M_{j0}}n_b -\stackrel{i}{M_{jk}}m^{(k)}_b
\end{eqnarray}
We can use following two relations to write $F^{IJ}$ in a form which is more useful:
\begin{eqnarray} 
    &&dW^{(ik)}\stackrel{\Delta}{=}-W^{(ij)}\wedge W^{(jk)} + \mathcal{R}_{abpq} m^p_{(i)} m^q_{(k)} ,\label{dW} \\
&&\tensor{R}{_{\underleftarrow{ab}}^c_d}\,\ell^d\stackrel{\Delta}{=}(d\tilde{\omega})_{\underleftarrow{ab}}\,\,\ell^{c}, \label{Riemann tensor condition}
\end{eqnarray}
where $\mathcal{R}_{abcd}$ is the Riemann tensor on the horizon cross-sections. These lead to:
the curvature $F_{IJ} =dA_{IJ}+A_{IK}\wedge\, A^{K}{}_{J}$ corresponding
to this connection $A_{IJ}$
\begin{equation}\label{curvature}
     {F_{\underleftarrow{c d}}}^{IJ}\= -2 (d\tilde{\omega})_{cd} \, l^{[I}n^{J]}+2 \{dV^{(k)}+(\omega \wedge V^{(k)})+(V^{(i)}\wedge W^{(ik)})\}_{cd}\,  l^{[I}m^{J]}_{(k)} + \tensor{\mathcal{R}}{_{kl}^{ij}} m_c^{(k)}m_d^{(l)}m^I_{(i)}m^J_{(j)}.
\end{equation}
In the following, we shall use this expression \eqref{curvature} to determine action and quantities on the phase- space. 
But first, let us establish the zeroth law.\\

\emph{Zeroth Law:} From the fourth boundary  condition, it follows that:
\begin{equation}\label{wih}
    \lie_{\ell}\, \omega\=0=d\,\kappa_{(\ell)},
\end{equation}
where we have used the equation eqn.\eqref{omega_expansion}, and $\kappa_{(\ell)}$ is the acceleration of the vector field $\ell^{a}$ on the horizon. This proves the zeroth law: \emph{surface gravity is a constant on the horizon $\Delta$}.  We see that zeroth law emerges directly from the geometrical structure of the horizon. \\

The curvature of the connection component (rotation one form) $\omega$ will play an important role. From \eqref{Riemann tensor condition} it is easy to show that $d\omega$ is a purely spatial form and hence it can be written as:
\begin{equation}\label{domega}
    d\Tilde{\omega}\=C_{ij}\, m^{(i)}\wedge m^{(j)}
\end{equation}
where the scalars $C_{ij}$ are related to the Weyl tensor. The weak isolation condition eqn. \eqref{wih} implies that $d\lie_{\ell}\, \omega\=0$. Unlike in four dimensions, the components of the Weyl tensors are not automatically Lie-dragged along the horizon generator. In fact, their behavior depends on the choice of $m^{(i)}$. For example, if one imposes the requirement that $\lie _{\l}m_{a}^{(i)}\=0$, then $\lie_\ell C_{ij}\=0$ is true by construction. Even without this requirement though, and with a suitable choice of basis vector fields on the horizon, one may impose the requirement that the Weyl scalars are Lie-dragged along the horizon generator.  This is physically acceptable since the condition of no flux crossing the horizon implies that the Weyl tensors associated with the foliation must also remain independent of the $v$ coordinate. On the horizon  $\lie_{\ell} m^{(i)}_{\underleftarrow{b}}\=-\stackrel{i}{M_{j0}}m^{(j)}_{\underleftarrow{b}}$. In the following, we set $\lie_\ell C_{ij}\=0$, but the basis vectors $m_{a}^{(i)}$ are not lie dragged along $\ell^{a}$. This puts the following constraints on the $\stackrel{i}{M}_{j0}$ :
\begin{equation}\label{constraint 1}
   \begin{array}{c @{\quad\text{or}\quad} c}
 C_{ik} \stackrel{k}{M_{j0}}-\stackrel{i}{M_{k0}} C_{kj}\=0 & [C,M_0]\=0
\end{array}
\end{equation}
Similarly if $\phi^a$ is the vector field corresponding to rotational symmetry then $\lie_{\phi}\, \omega \= 0$. If we assume $\lie_{\phi}\, m^{(i)}\= {A^i}_j m^{(j)}$ for some ${A}_{ij}$ antisymmetric in $i$ and $j$, then $\lie_\phi C_{ij}\=0$, will give rise to the following constraint:
\begin{equation}\label{constraint 2}
\begin{array}{c @{\quad\text{or}\quad} c}
 C_{ik} {A^k}_j-{A^i}_k C_{kj}\=0,  & [C,A]\=0.
\end{array}
\end{equation}
%
\emph{First law:}
The Lagrangian $5$-form for EGB theory in $5$ dimensions is given by:
\begin{equation}\label{Lagrangian}
   (16\pi \,G )\,  L= \{\Sigma_{IJ}+\alpha_2\,\Sigma_{IJKL}\wedge F^{KL}\}\wedge F^{IJ}+d\,\left(\, \Tilde{\Sigma}_{IJ}\wedge A^{IJ}\, \right),
\end{equation}
where, $F_{IJ}$ is the curvature defined before and 
\begin{equation}
    \Tilde{\Sigma}_{IJ}=\Sigma_{IJ}+2\alpha_2 \,\Sigma_{IJKL}\wedge F^{KL},
\end{equation}
with $\Sigma_{IJ}=(1/3!)\, \epsilon_{IJKLM}\, e^K\wedge e^L \wedge e^M$ and $\Sigma_{IJKL}=\epsilon_{IJKLM}\, e^M$, and $\alpha_2$ is the Gauss- Bonnet coupling constant. The boundary term is analogous to the Gibbons-Hawking term in GR and thus the variation of the Lagrangian becomes simpler to implement. The expression for  $\Tilde{\Sigma}_{IJ}$ on horizon in terms of internal vectors can be written as:
\begin{equation}\label{Sigma}
    \underleftarrow{\Tilde{\Sigma}}_{IJ} \= 2\,^3\epsilon(1+2\alpha_2 \mathcal{R})\, \ell_{[I}n_{J]} +8\alpha_2 \,^3\epsilon\, C_{ij} m_I^{(i)} m_J^{(j)} + 2\beta_{(k)} \ell_{[I} m_{J]}^{(k)},
\end{equation}
where $\mathcal{R}$ is the scalar curvature of the horizon cross-sections. The variation with respect to the tetrads ($e$) gives the following equation of motion:
\begin{equation}
    \Sigma_{IJM}\, F^{IJ}+\alpha_2 \left(\epsilon_{IJKLM} F^{IJ} \wedge F^{KL}\right)=0\, ,
\end{equation}
where $\Sigma_{IJM}=(1/2)~\epsilon_{IJKLM}\, e^{K}\wedge e^{L}$. The variation with respect to $A_{IJ}$ leads to:
\begin{equation}
    D\,\Tilde{\Sigma}_{IJ}=0,
\end{equation}
where $D$ is the gauge covariant derivative operator, $D\lambda^{I}=\partial\lambda^{I}+A^{I}{}_{J}\, \lambda^{J}$, for any internal vector $\lambda^{I}$. Apart from these equations, one also gets boundary terms which has to vanish for the action principle to be well- defined. The variation of the action leads to the following integral:
\begin{equation}
 \delta S=   \int_{\partial M=\Delta\,\cup\, i^{0}} \,\delta\Tilde{\Sigma}_{IJ} \wedge  A^{IJ}.
\end{equation}
The quantities at the asymptotic boundary vanish by appropriate fall- off conditions on the fields at $i^{0}$, whereas those on the horizon need to be properly evaluated. We now show that this integral vanishes on $\Delta$. Using the expressions of $A_{IJ}$ from eqn. \eqref{connection_expansion},  $F_{IJ}$ from \eqref{curvature}, and $ e_{a}^I\=-n_{a} \ell^I + m_{a}^{(i)} m_{(i)}^I$, we get
\begin{equation}\label{integrand}
     (\delta\Tilde{\Sigma}_{IJ}\wedge  A^{IJ})\=2\,\delta\,[{}^{3}\epsilon\,(1+2\alpha_2 \mathcal{R})]\wedge  \omega +8 \alpha_2 [\,\delta(^3 \epsilon C_{ij})\wedge W^{(ij)}]
\end{equation}
From the variational principle, the co-tetrad fields are fixed on the initial cross section $\mathbb S_{i}$, so $\delta m_b^{(i)}\=0$, on the initial cross section, which in turn fixes the volume form, i.e. $\delta ^3\epsilon\=0$ on $\mathbb{S}_{i}$ . Since the volume form is Lie dragged along the horizon generator, and $\delta \ell=c\ell$, $\lie_\ell (\delta ^3 \epsilon)\=0$ throughout the horizon. Further, fixing $\omega$ on the initial cross section $\mathbb{S}_{i}$ fixes $C_{ij}$ and since $\lie_\ell C_{ij}\=0$, this again implies $\lie_{\ell} (\delta C_{ij})\=0$. By fixing $W^{(ij)}$ on the initial cross section the Riemann curvature tensor intrinsic to the horizon cross section, $\mathcal{R}_{abcd}$ gets fixed, which ensures that the Ricci scalar $\mathcal{R}$ is fixed on $\mathbb{S}_{\i}$. Since $\mathcal{R}$ is Lie-dragged along $\ell$, $\delta \mathcal{R}$ is also Lie dragged along $\ell$. Once these quantities are fixed on $\mathbb{S}_{i}$ they are fixed on the $\Delta$ and hence this proves that the surface term is zero due to the boundary conditions on the horizon and the action principle is well defined, therefore equations of motion follow from $\delta S=0$.\\

\begin{figure}
    \centering
    \includegraphics[width=0.45\linewidth]{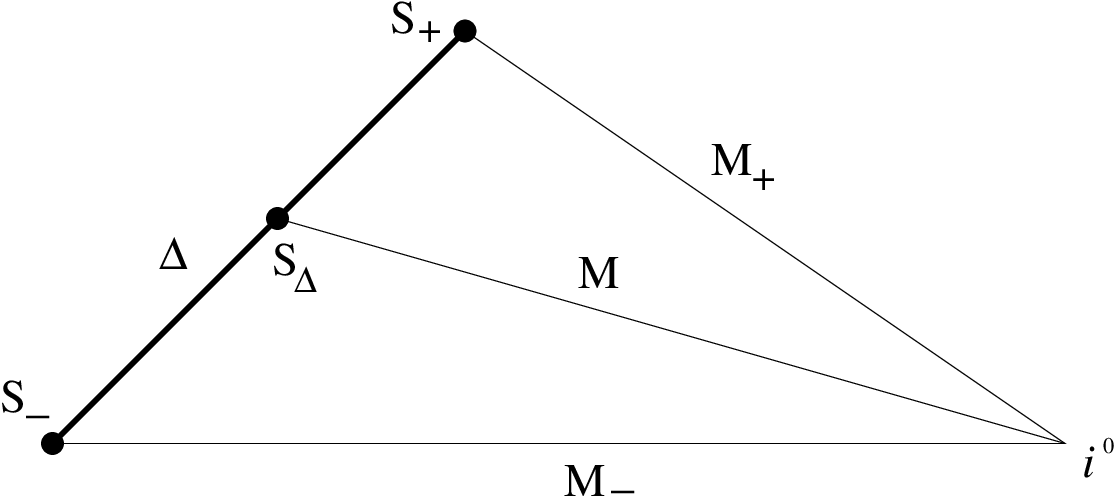}
    \caption{$M_{\pm}$ are two partial Cauchy surfaces enclosing a region of space-time and intersecting $\Delta$ in the 3-surface $S_{\pm}$ respectively,and extend to spatial infinity $i^0$. Another Cauchy slice M is drawn which intersects $\Delta$ in $S_{\Delta}$}
    \label{fig1}
\end{figure}

To obtain the symplectic structure, we proceed as follows. The variation of the Lagrangian $5$- form (in $5$- dimensions) gives the equation of motion and a boundary term: $\delta L=(EOM)\, \delta (fields) +d \theta(\delta)$, with $\theta(\delta)$ being the symplectic potential, a  spacetime $4$ form, but a $1$- form on phase space. The symplectic current is a closed $2$- form on phase space, obtained from the antisymmetric combination of symplectic potential: $J(\delta_{1}, \delta_{2})=\delta_{1}\, \theta(\delta_{2})-\delta_{2}\, \theta(\delta_{1})$. From the Lagrangian in \eqref{Lagrangian}, the symplectic potential is   $\theta(\delta)=\delta \Tilde{\Sigma}_{IJ}\wedge A^{IJ}$. The symplectic current can be written as:
 \begin{equation}\label{symplectic_current}
16 \pi G \,J(\delta_{1}, \delta_{2})=- (\delta_1 \Tilde{\Sigma}_{IJ} \wedge \delta_2 A^{IJ} - \delta_2\Tilde{\Sigma}_{IJ} \wedge \delta_1 A^{IJ}).
\end{equation}
Let the manifold $\mathcal{M}$ be such that it is bounded by the horizon $\Delta$, two partial Cauchy slices $M_{+}$ and $M_{-}$, and the spatial infinity $i^{0}$, (see fig. \ref{fig1}). Since $J(\delta_{1}, \delta_{2})$ is closed, on $\mathcal{M}=\Delta\cup M_{+}\cup M_{-}\cup i^{0}$, we get:
\begin{eqnarray}
\left[\, \int_{M+} -\int_{M_{-}} -\int_{\Delta} \,\right] \, J(\delta_{1}, \delta_{2})= 0,
\end{eqnarray}
where the quantity on $i^{0}$ is taken to vanish by choice of appropriate fall- off conditions on the fields. 
On $M_{+}$ or $M_{-}$, the value of $J(\delta_{1}, \delta_{2})$ is as given in \eqref{symplectic_current}, but on $\Delta$, the expression simplifies:
\begin{eqnarray}
  16 \pi G  J(\delta_{1}, \delta_{2})|_{\Delta}&=& -2\int_{\Delta}\delta_{1}\{^3 \epsilon (1+2\alpha_2 \mathcal{R})\} \wedge \delta_2 \omega- (1\leftrightarrow 2) -8\alpha_2 \int_{\Delta} \{\delta_{1}(^3\epsilon C_{ij})\wedge\delta_2(W^{(ij)})- (1\leftrightarrow 2) \nonumber\\
   =-&2&\int_{\Delta}\delta_{1}\{^3 \epsilon (1+2\alpha_2 \mathcal{R})\} \wedge \delta_2(-\kappa n)- (1\leftrightarrow 2) -8\alpha_2 \int_{\Delta} \{\delta_{1}(^3\epsilon C_{ij})\wedge\delta_2(- \stackrel{i}{M_{j0}} n)- (1\leftrightarrow 2)
\end{eqnarray}
where in the second line of the above equation we have used \eqref{omega_expansion} and \eqref{W and V}. Now we define following potentials for the Ricci rotation coefficients appearing in above equation:
\begin{equation}
\begin{array}{c @{\quad\text{and}\quad} c}
    \lie_\ell \psi\=\kappa_{(l)} & \lie_\ell \psi^{ij}\=\stackrel{i}{M_{j0}}
    \end{array}
\end{equation}
such that $\psi, \psi^{ij}$ vanish on $S_-$. Now we can use these potentials and the fact that $dC_{ij}=-[C,M_0]_{ij}\,n+ \alpha_{ijk} m^{(k)}$ which follows from \eqref{domega} and \eqref{constraint 1}, to show that:
\begin{eqnarray}
16\pi G   \,  \Omega(\delta_1, \delta_2)&=&\left\{-\int_M  \left[\delta_1 \Tilde{\Sigma}_{IJ} \wedge \delta_2 A^{IJ} \right] - 2 \oint_{S_{\Delta}}\left[\delta_1(^3 \epsilon (1+2\alpha_2 \mathcal{R}))\delta_2 \psi \right] \nonumber - 8\alpha_2 \oint_{S_{\Delta}}\left[\delta_1(^3 \epsilon\, C_{ij})\, \delta_{2} \,\psi^{ij} \right]\right\} -(1\leftrightarrow 2).
\end{eqnarray}
This symplectic structure is independent of the choice of hypersurface.\\

To determine the first law for an axially symmetric horizon, we introduce the vector field on the spacetime  $t^{a}=B_{(\ell,t)}\ell^a-\Omega_{t} \,\phi^{a}$, where $\phi^{a}$ is the angular Killing vector field corresponding to rotational symmetry of the cross- section and $B_{(\ell,t)}$, $\Omega_t$ are some constants on $\Delta$. More precisely we assume that the following conditions hold true: $\lie_{\phi}\, \omega\= 0$, $\lie_{\phi}\, q_{ab}\=0$, and $\lie_{\ell}\, \phi \=0$. Then we define $16 \pi G\,X_t(\delta):=\Omega(\delta,\delta_t)$, which using the equations of motion can be expressed as  as follows:
\begin{equation}
    16 \pi G\,X_t(\delta)=\oint_{\partial M} \delta \Tilde{\Sigma}^{IJ}(t\cdot A)_{IJ}+(t\cdot\Tilde{\Sigma}^{IJ})\wedge \delta A_{IJ}-\left\{2\oint_{S_{\Delta}}\delta\{^3 \epsilon (1+2\alpha_2 \mathcal{R})\}\delta_{t} \psi+ 8\alpha_2 \oint_{S_{\Delta}}\delta(^3 \epsilon C_{ij}) \delta_{t} \psi^{ij}\right\} -(\delta \leftrightarrow \delta_{t})
\end{equation}
Using \eqref{connection_expansion} and \eqref{Sigma}, after a calculations of few lines, we get;
\begin{equation}
    X_{t}(\delta)= \frac{1}{8 \pi G}\,\left[\oint_{S_{\Delta}}- \oint_{S_{\infty}}\right] \, \delta\{ {}^3 \epsilon(1+2\alpha \mathcal{R})\}\,\kappa_{t} - \Omega_{t}\, \delta[\phi\cdot \omega(1+2\alpha_2\mathcal{R})\,^3\epsilon+4\alpha_2(\phi\cdot W^{(ij)})\,{}^3\epsilon \, C_{ij}] +\delta E^{t}_{\infty}.
\end{equation}
Here, $E^{t}_{\infty}$ is the contribution to the ADM energy. In order to show that $\delta_{t} $ is Hamiltonian, we must show that $X_{t}(\delta)=\delta H_{t}$ is a closed form. This requires that the surface gravity $\kappa_t$ and angular velocity $\Omega_{t}$ must be function of area and angular momentum respectively. Naturally, this may be written in the following form:
\begin{equation}\label{first_law}
    \delta E^{t}_{\Delta}=\frac{1}{8\pi G}\,\kappa_{t}\, \delta\Tilde{a}_{\Delta}+\Omega_{t}\, \delta J_{\Delta},
\end{equation}
where, the quantities are defined as follows: $H^{t}=E^{t}_{\infty}-E^{t}_{\Delta}$. The quantity behaving like the area $\tilde{a}_{\Delta}$ is given by: 
\begin{equation}
    \Tilde{a}_{\Delta}=\oint_{S_{\Delta}}\, {}^3 \epsilon\, (1+2\alpha_2 \mathcal{R}) ,
\end{equation}
whereas the angular momentum $J_{\Delta}$ is given by the following expression:
\begin{equation}\label{angular momentum}
    J_{\Delta}=-\frac{1}{8\pi G}\oint_{S_{\Delta}}[\phi\cdot \omega \, (1+2\alpha_2\mathcal{R})\,^3\epsilon+4\alpha_2(\phi\cdot W^{(ij)})\,{}^3\epsilon \, C_{ij}].
\end{equation}
The requirement that $X_{t}(\delta)$ be Hamiltonian further implies that the following condition hold true:
\begin{equation}
    \frac{\partial \kappa_t(\Tilde{a}_{\Delta},J_{\Delta})}{\partial J_{\Delta}}= 8\pi G\frac{\partial \Omega_t(\Tilde{a}_{\Delta},J_{\Delta})}{\partial \Tilde{a_{\Delta}}}.
\end{equation}
These along with the above constraint ensure that the one form $X_t(\delta)$ is closed.  It is also not difficult to show that the angular momentum is conserved,  $\lie_{t} J_{\Delta}=0$.\\

In this paper, we have shown that the first law of black hole mechanics for rotating horizons in the five dimensional Einstein- Gauss- Bonnet holds true for rotating black holes as well. The first law of black hole mechanics given above, in eqn. \eqref{first_law} is in the
differential form of the first law of thermodynamics. The quantity representing the entropy, is not the horizon area, but is modified
due to contribution from the Gauss- Bonnet part of action. This is expected since such a term also arises due to Wald-Iyer \cite{Wald:1993nt,Iyer:1994ys} definition of entropy as a Noether charge. The important contribution in this paper is the expression of angular momentum in \eqref{angular momentum}, which has been obtained in terms of geometric quantities on the horizon and the Ricci scalar. Although no exact rotating black hole solution is known in the five dimensional Gauss- Bonnet theory, the formalism of IH allows the extraction of such a quantity defined on the horizon. Furthermore, in the limit of the GB coupling $\alpha_2=0$, the expression of $J_{\Delta}$ reduces to that in GR. Of course the value of angular momentum remains ambiguous upto a total variation. The extension of this calculation to $N$-dimensional Lovelock action is also possible and shall be discussed in an upcoming paper. The first law for 
isolated horizons in this theory also takes the standard expression eqn. \eqref{first_law}. The modified expressions for area and angular momentum take the following forms:
\begin{equation}
    \Tilde{a}_{\Delta}=\oint_{S_{\Delta}}~ ^{N-2} \epsilon~\mathcal{L},
\end{equation}
\begin{equation}
   J_{\Delta}=-\frac{1}{8 \pi G}\oint_{S_{\Delta}} [(\phi\cdot \omega) ~\mathcal{L}+(\phi \cdot W^{(kl)})~\Tilde {C}_{kl}]~^{N-2} \epsilon~ ,
\end{equation} 
where $\mathcal{L}$ is the Lanczos-Lovelock scalar of the horizon. The expressions for Lanczos-Lovelock scalar $\mathcal{L}$ and $\Tilde{C}^{kl}$ are:
\begin{equation}
    \mathcal{L}=\sum_{p=1}^{\bigl[\frac{N-1}{2}\bigr]} p \alpha_p\left(\frac{1}{2^{p-1}}\right)
\delta^{i_1 i_2 \cdots i_{2p-3} i_{2p-2}}_{j_1 j_2 \cdots j_{2p-3} j_{2p-2}}
{\mathcal{R}_{i_1 i_2}}^{j_1 j_2}\cdots
{\mathcal{R}_{i_{2p-3}\, i_{2p-2}}}^{j_{2p-3} j_{2p-2}},
\end{equation}
\begin{equation}
    \Tilde{C}_{kl}=\sum_{p=1}^{\left[\frac{N-1}{2}\right]} p\alpha_p
\left(\frac{p-1}{2^{p-2}}\right)
\delta^{ij\,i_1 i_2 \cdots i_{2p-5}\, i_{2p-4}}_{kl\,j_1 j_2 \cdots j_{2p-5}\,j_{2p-4}}
{\mathcal{R}_{i_1 i_2}}^{j_1 j_2}\cdots
{\mathcal{R}_{i_{2p-5}\, i_{2p-4}}}^{j_{2p-5} j_{2p-4}}
\,C_{ij} .
\end{equation}
This calculation requires stronger boundary conditions 
than used in this paper, and the details shall be discussed elsewhere.\\

\textbf{Acknowledgements:} AC acknowledges the support of  DAE-BRNS project $58/14/25/2019$-BRNS. SD acknowledges the financial support provided by DST vide Grant No. DST/INSPIRE Fellowship/2020/IF200537.


\end{document}